\documentclass[journal,10pt]{IEEEtran}
\IEEEoverridecommandlockouts
\usepackage{}
\usepackage{amsmath}
\usepackage{graphicx}
\usepackage{amssymb}
\usepackage{amsmath,amsthm} % Set the style of Theorem
\usepackage{float}
\usepackage{enumerate}
\usepackage{epstopdf}
\usepackage{dblfloatfix}
\usepackage{caption}
\usepackage{amsfonts}
\usepackage{fancyhdr}
\usepackage{bbm}
\usepackage{cases}
\usepackage{cite}
\usepackage{breqn}
\usepackage{multirow}
\usepackage{mathrsfs}
\usepackage{algpascal}
\usepackage{algc}
\usepackage{algorithmicx}
\usepackage[ruled]{algorithm}
\usepackage{algpseudocode}
\usepackage{graphics}
\usepackage{epsfig}
\usepackage{mathrsfs} % the form of set
\usepackage{dsfont}
\usepackage{color}
\usepackage{hyperref}
\usepackage{hyperref}

\floatname{algorithm}{Algorithm}

\algdef{SE}[DOWHILE]{Do}{doWhile}{\algorithmicdo}[1]{\algorithmicwhile\ #1}%

\ifCLASSINFOpdf
 \else
\fi

\usepackage{amsmath}
\usepackage{algorithm}
\usepackage{algpseudocode}

\begin{document}
\theoremstyle{definition} % Set the style of Theorem
\newtheorem{theorem}{Theorem}[section]
\newtheorem{definition}[theorem]{Definition}
\newtheorem{lemma}[theorem]{Lemma}
\newtheorem{example}[theorem]{Example}
\newtheorem{Proposition}[theorem]{Proposition}
\newtheorem{Corollary}[theorem]{Corollary}
%\newtheorem{remark}[theorem]{Remark}

% \title{Deep Learning and Transfer Learning-Inspired Beam Prediction In Vehicle Communications}
% \title{A Low-overhead Deep Transfer Learning Approach for Beam Prediction in Vehicle Communications}
\title{A Deep Transfer Learning-Based Low-overhead Beam Prediction in Vehicle Communications}

\author{Zhiqiang Xiao, Yuwen Cao\textsuperscript{},~\IEEEmembership{}Mondher Bouazizi\textsuperscript{},~\IEEEmembership{}Tomoaki Ohtsuki\textsuperscript{},~\IEEEmembership{Senior Member, IEEE}, and Shahid Mumtaz\textsuperscript{},~\IEEEmembership{Senior Member, IEEE}\vspace{-0.45cm}
\thanks{
This work was supported in part by JST ASPIRE Grant Number JPMJAP2326, Japan. 
Z. Xiao and Y. Cao are with the
College of Information Science and Technology, Donghua University, Shanghai, China. M. Bouazizi and T. Ohtsuki are with the Department of Information and Computer Science, Keio University, Yokohama, Japan. S. Mumtaz is with the Department of Engineering,
Nottingham Trent University, UK.}
}

% \markboth{IEEE WIRELESS COMMUNICATIONS LETTERS,~Vol.~XX, No.~XX, XXX~2025}
{}

\maketitle

\begin{abstract}
%Transfer learning has been widely applied in beamforming for vehicle communications to enable models to rapidly adapt to new environments. However, existing approaches primarily rely on simple fine-tuning, when there is a significant difference in data distribution between the target domain and the source domain, simple fine-tuning limits the model's performance in the target domain, resulting in insufficient accuracy in beam prediction. To tackle this problem, we propose a transfer learning based beam prediciton method that combines fine-tuning with domain adaptation. First, we train the model on the source domain to obtain a pre-trained model. Next, we integrate a domain classifier into fine-tuning the pre-trained model. Using the pre-trained model, we extract high-level domain features from both source and target domain data. In adversarial training with the domain classifier, the model extracts domain-invariant features. By jointly fine-tuning the pre-trained model and training the domain classifier, we further enhance the prediction accuracy of transfer learning.  Simulation results demonstrate that the proposed transfer learning method achieves an effective achievable rate close to that of training from scratch in the target domain and the scenario with known optimal beamforming, while significantly reducing beam training overhead.
Existing transfer learning-based beam prediction approaches primarily rely on simple fine-tuning. When there is a significant difference in data distribution between the target domain and the source domain, simple fine-tuning limits the model's performance in the target domain. To tackle this problem, we propose a transfer learning-based beam prediction method that combines fine-tuning with domain adaptation. We integrate a domain classifier into fine-tuning the pre-trained model. The model extracts domain-invariant features in adversarial training with domain classifier, which can enhance model performance in the target domain. Simulation results demonstrate that the proposed transfer learning-based beam prediction method achieves better achievable rate performance than the pure fine-tuning method in the target domain, and close to those when the training is done from scratch on the target domain.
\end{abstract}


\section{Introduction}
Research on vehicle communications has achieved remarkable progress in recent years. In addition, cutting-edge technologies such as modulation techniques, intelligent network optimization and deep learning are driving the advancement of vehicle communications\cite{10381876,11145277,10422712,9463461}. 
% \textcolor{red}{In the context of 6G, the key technologies such as machine learning and RIS (Reconfigurable Intelligent Surface) have effectively enhanced the performance metrics of beam management and beam prediction\cite{10422712}.} 
In this context, for accurate beam prediction in 
millimeter-wave (mmWave) communications, existing machine learning-based beam training approaches require massive data and time to retrain the model from scratch, leading to unacceptable training overhead and data transmission delay. 

In recent research, transfer learning has been utilized to reduce the time and overhead required for training models to predict beams. Works such as \cite{10333610,10900958,10170792} leverage transfer learning to address specific challenges in their respective domains, e.g., defect detection, beam tracking, and beam selection. They highlight the benefits of transfer learning in reducing training time, improving accuracy, and adapting to dynamic or unseen environments. 
%More concretely, the authors in \cite{10333610} address the challenge of frequent beam training in dynamic environments by proposing a transfer learning-based beam training method for intelligent omni-surfaces. Laskos \textit{et al.} in \cite{10900958} explore the use of transfer learning to accelerate the learning process for mmWave beam tracking in mobile environments. 
%Reference \cite{10279177} proposes a framework that focuses on adapting pre-trained models to unseen environments in vehicle-to-everything (V2X) mmWave communication.
Besides, \cite{10762896,10463115} apply transfer learning to beam prediction tasks, aiming to improve model performance in new environments with minimal additional training data. 
They highlight the efficiency and adaptability of the transfer learning strategy and tackle data privacy concerns, thereby reducing the amount of required training data. The authors  in \cite{xia2024intelligentanglemapbasedbeam} use only location information to achieve fast and accurate beam alignment for both line-of-sight (LoS) and non-line-of-sight (NLoS) users in reconfigurable intelligent surface (RIS)-aided mmWave systems without beam scanning. 
%Currently, transfer learning methods based on fine-tuning have become increasingly mature. In \cite{10900958,10170792}, the authors develop a beam prediction model that is pre-trained offline and then fine-tuned online using a small dataset from the new environment. The model reduces training overhead and improves convergence speed, achieving near-optimal performance in dynamic scenarios. 
However, current transfer learning methods for beamforming primarily rely on simple fine-tuning, when there is a significant difference in data distribution between the source and target domains, pure fine-tuning leads to degraded model performance in the target domain and limited generalization capability.

Domain adaptation has been demonstrated in previous research to play a significant role in reducing the distribution discrepancy between the source and target domains and enhancing the model's generalization capability across different domains.
In \cite{10440352,10218991}, the authors address domain adaptation, focusing on adapting models from a source domain to a target domain with different distributions.
\cite{10218991} addresses the challenge of adapting deep neural networks (DNNs) to target domains without access to the source dataset. %It uses a teacher-student framework to progressively adapt class-wise feature centers from the source to the target domain. %generating compact target features and improving pseudo-label accuracy. 
Tu \textit{et al.} \cite{10867854} 
%and Elmaghbub \textit{et al.} \cite{10279347} 
leverage domain adaptation feature discrepancy minimization to achieve better performance in target domains particularly when dealing with limited labeled data or significant domain shifts.
%The authors in \cite{xia2024intelligentanglemapbasedbeam} present a semi-supervised domain adaptive synthetic aperture radar (SAR) ship detection method. The method incorporates the Mean Teacher model with a knowledge distillation framework to capture domain-invariant features and mitigate domain discrepancies.  
%In \cite{10279347}, authors introduce an unsupervised domain adaptation framework that separates device-specific features from domain-specific features, the method improves classification accuracy in both short-term and long-term temporal adaptations. 

The distribution discrepancy between the source and target domains can limit transfer learning's beam prediction performance in the target domain. In this letter, we study the interplay between domain adaptation and fine-tuning, followed by proposing a combined approach that can handle the distribution discrepancy between the source and target domains. We employ a network pre-trained on the source domain. We freeze the front layers of the network and conduct adversarial training between the domain classifier and the network to extract domain-invariant features from the source and target domains. We improve the beam prediction performance of the model in the target domain while reducing the overhead of model training.
The major contributions of this letter are summarized as follows:
\begin{itemize}
\item We propose a transfer learning-based beam prediction method that employs fine-tuning combined with a domain classifier. The domain classifier can learn domain-invariant features between the source and target domains in adversarial training, effectively enhancing the model's performance on the target domain.
\item In addition, we devise a convolutional neural network (CNN)-based beam prediction framework that can effectively extract temporal and spatial relationships and high-level features, thus improving the prediction accuracy.
\item We apply the proposed methods to vehicle communication systems, considering a scenario where multiple base stations (BSs) serve a single user. Simulation comparisons are conducted against conventional beam selection methods from scratch and pure fine-tuning approaches, demonstrating the superiority of our proposed methods.
\end{itemize}

{\it Notations}: Bold lower case letters represent vectors, while bold upper case letters denote matrices.
$\left( {\cdot} \right)^{T}$ and $\left( {\cdot} \right)^{-1}$ represent the transpose and inverse operations of a matrix, respectively. $\Re \{\mathbf{x} \}$ and $\Im \{ \mathbf{x} \}$ mean the element-wise real and imaginary parts of $\mathbf{x}$, respectively. $\otimes$ denotes Kronecker product. In addition, 
$\left| {\mathbf{X}} \right|$ is the determinant of $\mathbf{X}$.

\section{System Model}
\begin{figure}
    \centering
    \includegraphics[width=6cm]{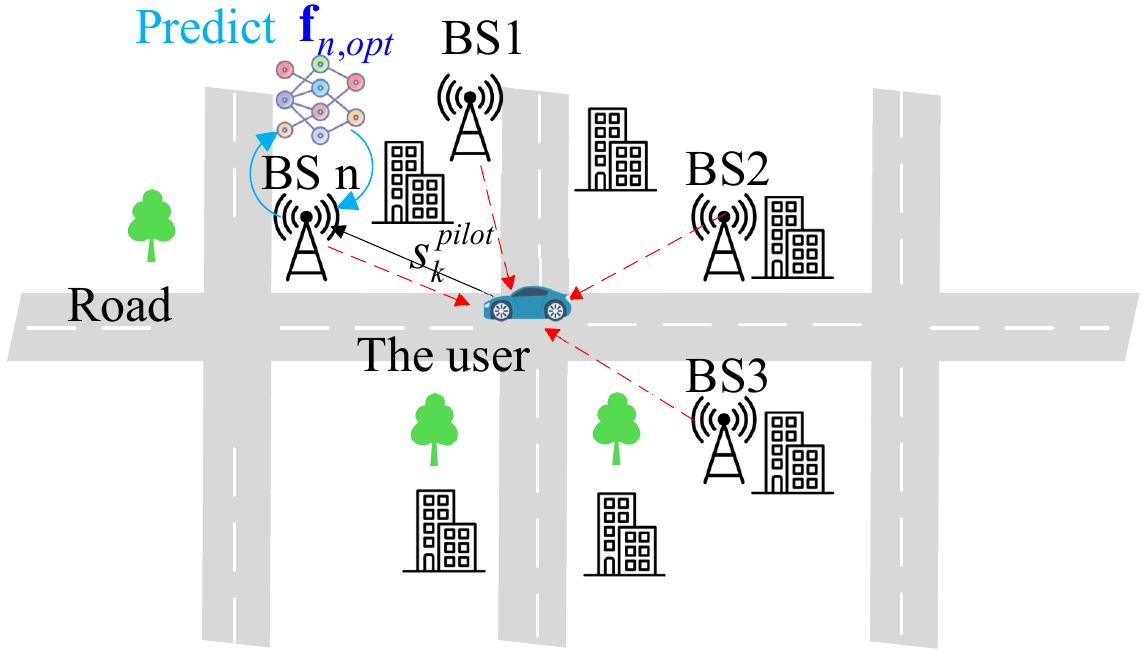}
    \caption{Multiple BSs simultaneously serve a user in the vehicle communication system.}
    \label{fig:1}
\end{figure}

Consider the vehicle communication system shown in \autoref{fig:1}, where multiple BSs simultaneously serve a vehicle, which can be considered as a user. For simplicity, each BS is equipped with 
$M$ antennas, while each vehicle is equipped with a single antenna.

We adopt the orthogonal frequency division multiplexing (OFDM) in our vehicle communication systems. Based on the mmWave channel model\cite{7990158}, the channel vector between the user and the $n$-th BS ${\mathbf{h}}_{k,n}$ at subcarrier $k$ can be expressed as 
\begin{equation}
\mathbf{h}_{k,n}= \sqrt{\frac{M}{L}} \sum_{l=1}^{L} \alpha_l \mathbf{a}_n(\phi_l, \theta_l) {e^{\frac{{ - j2\pi {\tau _l}k}}{K}}}{e^{{{ j2\pi f_{d,l}t}}}},
\end{equation}
where $L$ is the number of multipath, ${\alpha _l}$ is the complex gain of the $l$-th path, ${{\mathbf{a}}_n}({\phi _l},{\theta _l})$ is the array response vector of $n$-th BS, ${\tau_l}$ is delay of the $l$-th path.
Doppler shift in the $l$-th path ${f_{d,l}} = \frac{{v{f_c}}}{c}\cos {\theta _l}$, $f_c$ is carrier frequency, $v$ is relative velocity between transmitter and receiver. When the user sends pilot sequence $s_k^{pilot}$\cite{8395149}, the receive signal at $k$-th subcarrier and the $n$-th BS can be expressed as:
\begin{equation}
{r}_{k,n}^{(d)} = {\mathbf{f}}_{n,d}^T{{\mathbf{h}}_{k,n}}s_k^{pilot} + {\mathbf{f}}_{n,d}^T{{\mathbf{n}}_{k,n}},{\kern 1pt} \quad \,k = 1,2, \ldots ,K,   
\end{equation}
where ${\mathbf{n}}_{k,n}$ is the receive noise vector at $k$-th subcarrier and the $n$-th BS, and ${{\mathbf{f}}_d}$ is the $d$-th beamforming vector in the discrete Fourier transform (DFT) codebook used at the $n$-th BS.  
The achievable rate of the $n$-th BS in downlink can be expressed as:
\begin{equation}
R_n^{(d)} = \frac{1}{K}\sum\limits_{k = 1}^K {{{\log }_2}(1 + } \frac{{|{\mathbf{h}}_{k,n}^T{{\mathbf{f}}_d}{|^2}}}{{{\sigma ^2}}})
,\end{equation}
where ${\sigma ^2}$ denotes noise power. The DFT codebook is used at the BSs as 
${\mathbf{F}}$, which can be expressed as 
${\bf{F}} = {{\bf{F}}_{\rm{h}}} \otimes {{\bf{F}}_v}$. Herein, ${{\bf{F}}_h} = {[F_h^1,...,F_h^{{M_h}}]^T}$ and 
${{\mathbf{F}}_v} = {[F_v^1,...,F_v^{{M_v}}]^T}$ are the beamforming weight codebooks alongside the horizontal and vertical directions, respectively. Besides, $M_h$ and $M_v$ are the number of antenna elements in the horizontal and vertical axes, respectively. The optimal beam at the $n$-th BS in downlink ${\mathbf{f}}_{n,opt}$ can be expressed as:
\begin{equation}
\label{eq:004}
{{\mathbf{f}}_{n,opt}} = \mathop {\arg \max }\limits_{{{\mathbf{f}}_d} \in {\mathbf{F}}} \frac{1}{K}\sum\limits_{k = 1}^K {{{\log }_2}(1 + \frac{{|{\mathbf{h}}_{k,n}^T{{\mathbf{f}}_d}{|^2}}}{{{\sigma ^2}}}} )
.\end{equation}
We consider the scenario where multiple BSs serve a single user, with each BS selecting an optimal beam to maximize the achievable rate in downlink. The effective achievable rate in downlink $R_{{\rm{eff}}}$ can be obtained by
\begin{equation}
\label{eq:005}
\begin{split}
{R_{{\rm{eff}}}} = (1 - \frac{{{T_{{\rm{tr}}}}}}{{{T_{^{_B}}}}})\frac{1}{K}\sum\limits_{k = 1}^K {{{\log }_2}} \left( {1 + \frac{{{{\left| {\sum\limits_{n = 1}^{{N_{BS}}} {{\mathbf{h}}_{k,n}^T{{\mathbf{f}}_{n,opt}}} } \right|}^2}}}{{{\sigma ^2}}}} \right),
\end{split}
\end{equation}
where ${T_{tr}}$ represents the beam training time, ${{T_B}}$ corresponds to the beam coherence time, and ${{N_{BS}}}$ denotes the amount of active BSs. We individually predict the optimal beam for each BS. The formulation can illustrate the time overhead as the user moves. 
Notice that, we employ coordinated beamforming design to minimize mutual interference among BSs. At the user, 
assuming perfect frequency and carrier offset synchronization, the received signal is transformed into the frequency domain using fast Fourier transform (FFT). Besides, we utilize joint transmission in our system where coordinating BSs simultaneously deliver orthogonal frequency resources to the user to eliminate interference. Hence, the interference terms in the denominator of Eqs. (\ref{eq:004}) and (\ref{eq:005}) can be completely removed.

% there are no interference between each two users, i.e., the data to a single user to eliminate interference.}

As shown in \autoref{fig:1}, in the vehicle communication scenario, when the vehicle is in a different location and in range of different BSs which will serve it, this could be regarded as a new scenario that we treat as the target domain. The scenario with the original BSs distribution can be considered as the source domain. Due to differences in BS locations, the data distributions of received signals and achievable rates differ between the source and target domains. Transfer learning methods based solely on fine-tuning will consequently constrain the model's beam prediction performance in the target domain. 

\section{Proposed Low-overhead Beam Prediction Frameworks}
\subsection{Proposed Deep Learning Design}
\begin{figure}[tp]
    \begin{center}
    \includegraphics[width=8.5cm]{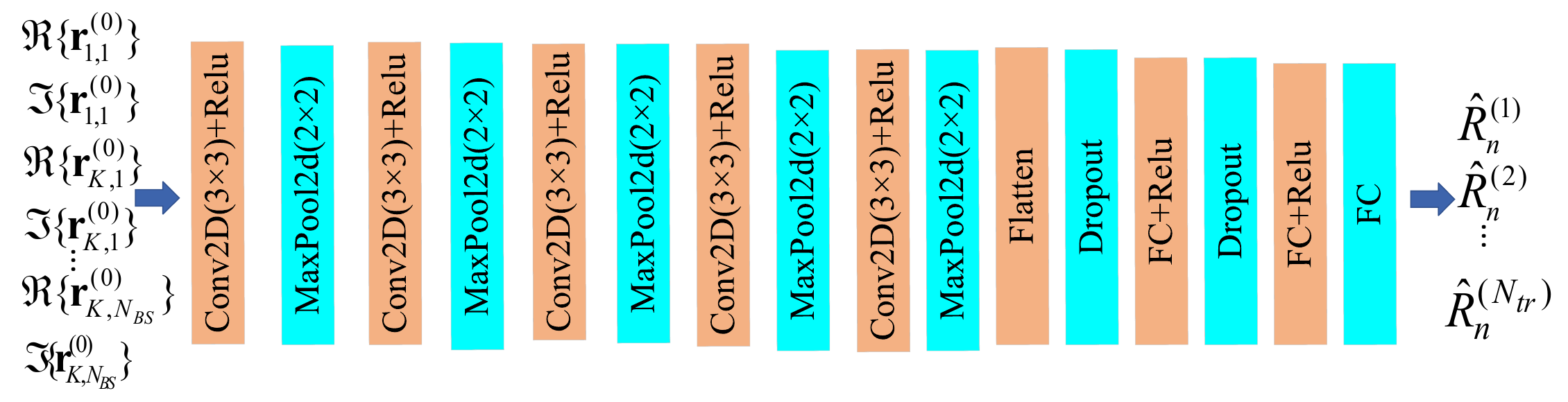}
    \end{center}
    \caption{\; An overview of the proposed CNN-based beam prediction framework, in which the first ${K_{DL}}$  samples of the $K$-point OFDM symbol are used as the input of the proposed network.} \label{fig:2}
\end{figure}

 We employ a CNN to predict the beamforming vector. The CNN model contains four convolutional layers. The CNN can achieve high accuracy and automatically extract higher-level features\cite{10816154}. In the uplink, users transmit pilot signals, and the received signals at the BS are in complex form. The CNN can effectively extract the temporal and spatial relationships and features contained in the received signals\cite{10816154,10146432,9314253}. The proposed CNN to predict beam in downlink is depicted in \autoref{fig:2}. The outputs of the network are the achievable rate in downlink for all beams in the codebook $\mathbf{F}$. We propose training a dedicated model for each BS to predict the optimal beam configuration specific to that station. 
In the uplink, we employ the first symbol $\mathbf{f}_0$ from the beamforming codebook $\mathbf{F}$ at the BS to obtain the received OFDM symbol, i.e.,
\setlength{\abovedisplayskip}{10pt} 
\setlength{\belowdisplayskip}{10pt} 
\begin{equation}
{r}_{k,n}^{(0)} = {\mathbf{f}}_0^T{{\mathbf{h}}_{k,n}}s_k^{pilot} + {\mathbf{f}}_0^T{{\mathbf{n}}_{k,n}},{\kern 1pt} \quad \,k = 1,2, \ldots ,K.
\end{equation}

For each data point, we set first ${K_{DL}}$  samples of the received $K$-point OFDM symbol, as the input of the proposed network.
We divide every received signal into two parts, 
$\Re \{ {r}_{k,n}^{(0)}\}$ and $\Im \{ {r}_{k,n}^{(0)}\}$, representing the real and imaginary components of $r_{k,n}^{(0)}$,
as the input of the CNN. The output of the model is the predicted achievable rate in downlink $\hat{R}_n^{(d)},d = 0,1,\ldots,{N_{tr}}-1$, when the $n$-th BS uses all beams in the codebook. The proposed deep learning method simultaneously predicts beamforming vectors for all BSs. Compared with the simple fully connected neural networks\cite{8395149}, the proposed CNN can extract the temporal and spatial relationships contained in the amplitude and phase of the received signal. We also propose different channel model, different receive signal as input of the model and different achievable rate as output, which can improve the model's ability to predict beams.

\subsection{Proposed Transfer Learning  method}
\begin{figure}[tp]
\setlength{\abovecaptionskip}{-0.1cm}
\setlength{\belowcaptionskip}{-0.5cm}
    \begin{center}
    \includegraphics[width=5.5cm]{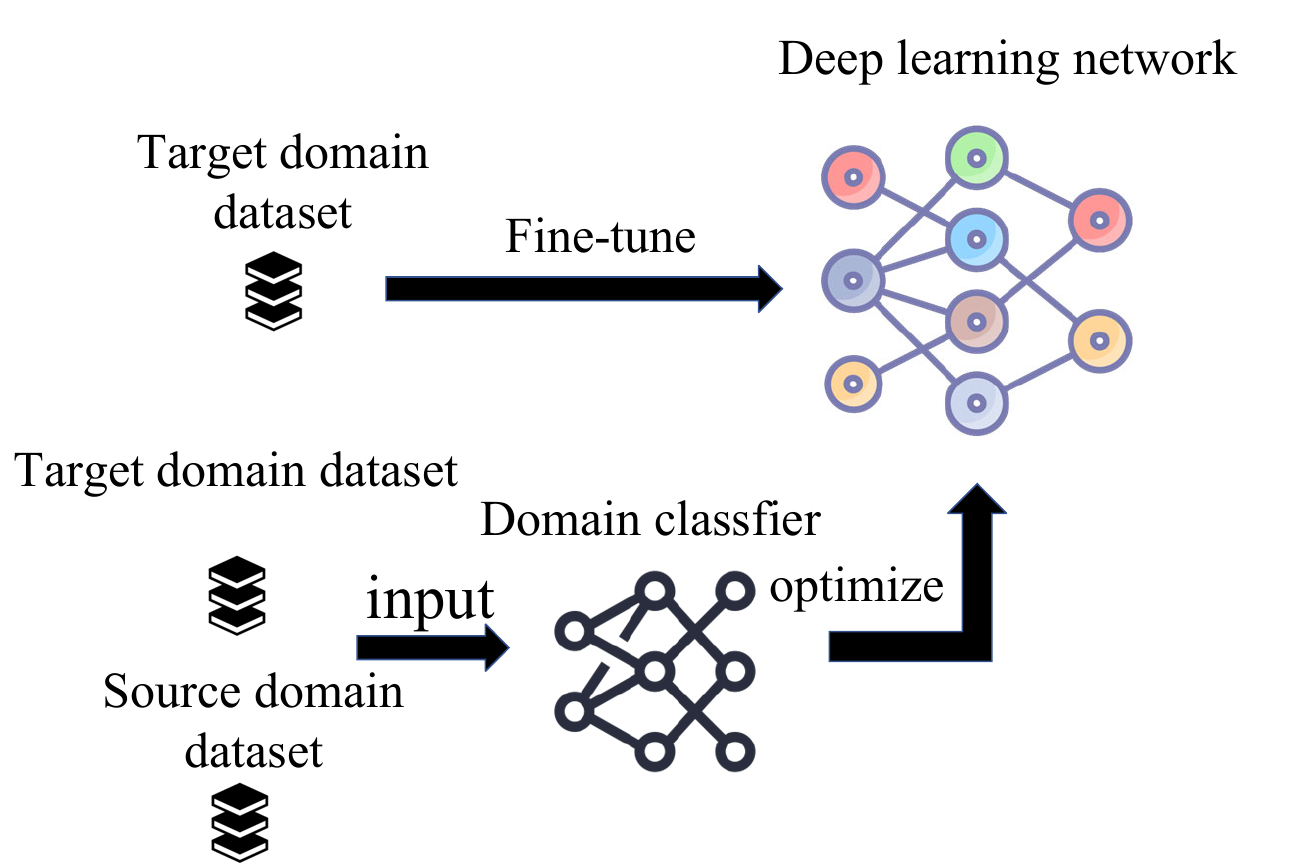}
    \end{center}
    \caption{\; Domain classifier and fine-tune architecture. We input both target domain and source domain dataset into the CNN to extract features, then input these features to the domain classifier network to predict domain label.} \label{fig:3}
\end{figure}

The transfer learning-based beam prediction method we propose combines fine-tuning with domain adaptation networks. The model trained on the source domain serves as the pre-trained model, as shown in \autoref{fig:3}.

\begin{figure}[tp]
\setlength{\abovecaptionskip}{-0.1cm}
\setlength{\belowcaptionskip}{-0.3cm}
    \begin{center}
    \includegraphics[width=5.5cm]{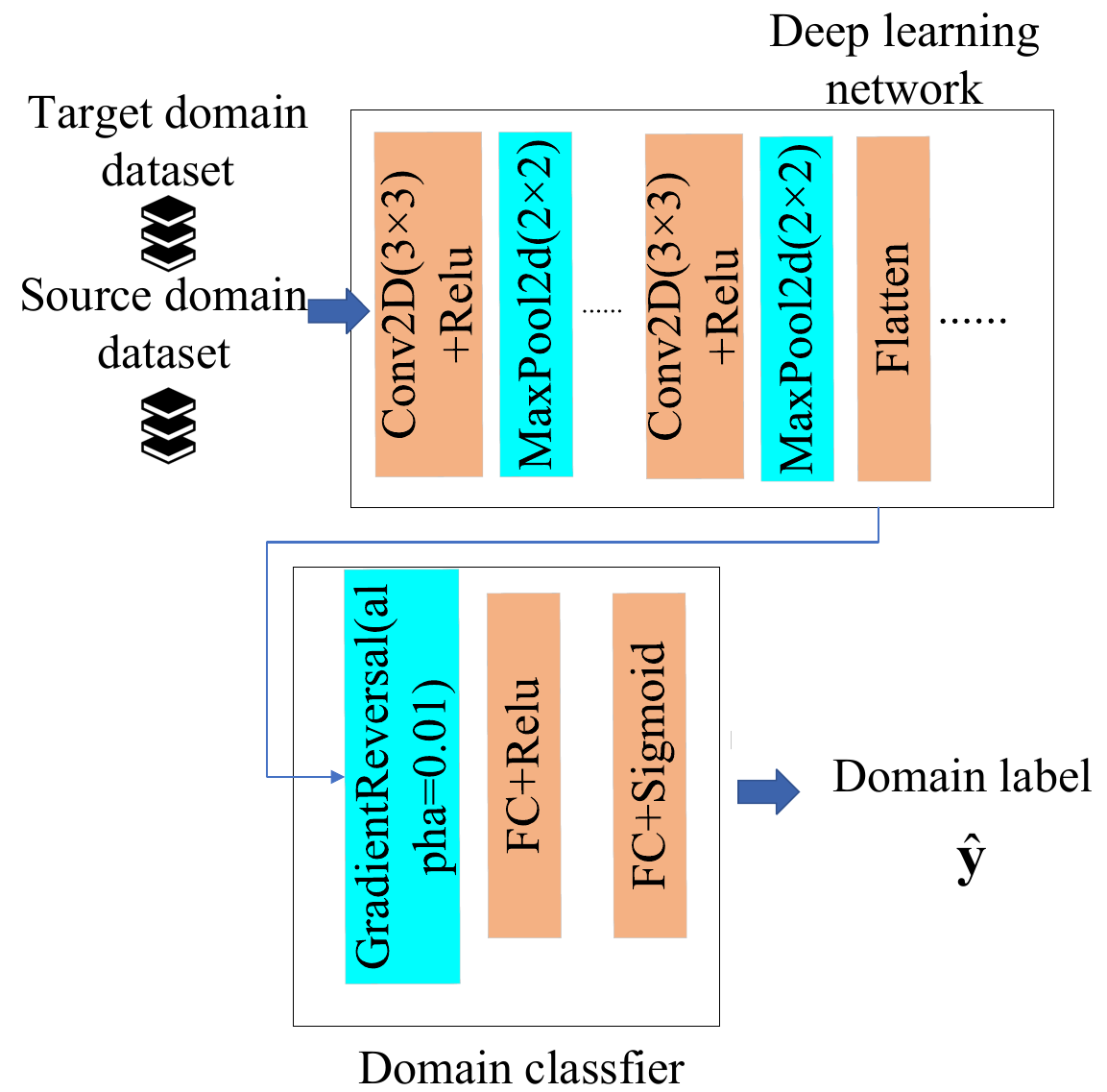}
    \end{center}
    \caption{\; Domain classifier network, the CNN extracts features as input.} \label{fig:4}
\end{figure}

Firstly, we perform fine-tuning. We freeze the first four layers of the pre-trained model as they serve as feature extraction layers and do not require further fine-tuning. We train the CNN using the target domain dataset. Additionally, we conduct domain adaptation (DA). We input both the source and target domain data into the pre-trained model, extract features using the pre-trained model, and use these features as input to the domain adaptation network. The domain adaptation network employs a domain classifier, as shown in \autoref{fig:4}. The domain classifier outputs the probability that the input data belongs to the source domain.

The loss function of the proposed CNN can be expressed as
\begin{equation}
\mathcal{L}(\theta ) = \sum\limits_{d = 0}^{{N_{tr}}-1} {MSE(R_n^{(d)},\hat R_n^{(d)})}  
,\end{equation}
where $R^{(d)}_n$ is the true achievable rate of the $n$-th BS when adopting the $d$-th beam, $\hat{R}^{(d)}_n$ is the predicted one, and $\theta$ are the parameters of the CNN.
The loss function of the domain classifier can be expressed as
\vspace{-0.2em} % 上移0.5行
\begin{equation}
\begin{split}
{\mathcal{L}_{\text{domain}}}({\theta _d}) = -\frac{1}{N}\sum\limits_{i = 1}^N {[{y_{_i}}\log ({{\hat p}_{_i}}} ) +\\ (1 - {y_{_i}})\log (1 - {{\hat p}_{_i}})],
\end{split}
\end{equation}
where ${y_{_i}}$ denotes the true domain label, ${{\hat p}_i}$ represents the predicted probability of belonging to the source domain. ${y_i} = 0$ indicates the target domain whereas ${y_i} = 1$ indicates source domain. ${\theta _d}$ denotes the parameters of domain classifier network. We optimize the whole model that includes the CNN and the domain classifier by minimizing the total loss function $% 
{\mathcal{L}_{{\rm{total}}}}(\theta ,{\theta _d})$. Thus, we have
\begin{equation}
{\mathcal{L}_{{\rm{total}}}}(\theta ,{\theta _d}) = \mathcal{L}(\theta ) + {\mathcal{L}_{\text{domain}}}({\theta _d})
.\end{equation}

In the domain classifier, the gradient reversal layer implicitly reverses the gradients of the domain classifier during backpropagation. This causes the feature extractor to implicitly maximize the domain classifier's loss while minimizing the loss of the CNN tasked with predicting the optimal beam. Through this adversarial training mechanism, the model is forced to learn high-level domain-invariant features, domain-invariant features encompassing both temporal and spatial characteristics are extracted, thereby enhancing the model's performance in the target domain, eliminating distribution discrepancies between the source and target domains and improving its performance on the target domain.
By utilizing the loss functions generated from fine-tuning and the domain classifier, we simultaneously optimize the model. Compared with existing transfer learning schemes, our proposed approach mitigates the inter-domain discrepancy to enhance the accuracy of beam prediction. However,  domain adaptation increases training overhead.


The pseudo-code for the proposed transfer learning-based beam prediction method is shown in Algorithm 1.

\begin{algorithm}[H]
\caption{Proposed Fine-tuning and domain adaptation beam prediction Framework}
\label{alg:fine_tune_model}
\begin{algorithmic}[1]
\Require Target domain data; source domain data; test data; training parameters 
\Ensure Transfer Learning models 
%\State \textbf{Reshape input data into 2D format}

\State \textbf{Initialize the models and data loaders}
\For{each BS}
        \State Load the pretrained deep learning model
        \State Freeze the model's first $freezed$ layers
    %\State Initialize optimizer, loss functions, and 
    %learning rate
    %\Statex \hspace{1em} scheduler

    \State \textbf{Train the model}
    \For{each epoch}
        \State Set the model to training mode
        \For{Target and source domain batches}
            \State Reshape the input data into 2D format
            \State Forward pass: Compute the model outputs and 
            \Statex \hspace{4em} the domain classifier outputs
            \State Compute the CNN loss 
                   ${\mathcal{L}}(\theta )$
            \State Compute ${\mathcal{L}_{\text{domain}}}({\theta _d}) $
            \State Total loss: 
            \Statex \hspace{4.5em} ${\mathcal{L}_{{\rm{total}}}}(\theta ,{\theta _d})  \leftarrow  {\mathcal{L}_{}}(\theta ) + {\mathcal{L}_{\text{domain}}}({\theta _d})$
            \State Backpropagate ${L_{{\rm{total}}}}(\theta ,{\theta _d})$ and update 
            \Statex \hspace{4.5em}parameters $\theta$, ${\theta _d}$
        \EndFor
    \EndFor

    \State \textbf{Save and store model}
\EndFor

\State \Return Transfer Learning models
\end{algorithmic}
\end{algorithm}
\section{Simulation Results}
\subsection{The Source Domain and Target Domain Scenario}
\begin{figure}[tp]
\setlength{\abovecaptionskip}{-0.2cm}
\setlength{\belowcaptionskip}{-0.5cm}
    \begin{center}
    \includegraphics[width=7cm]{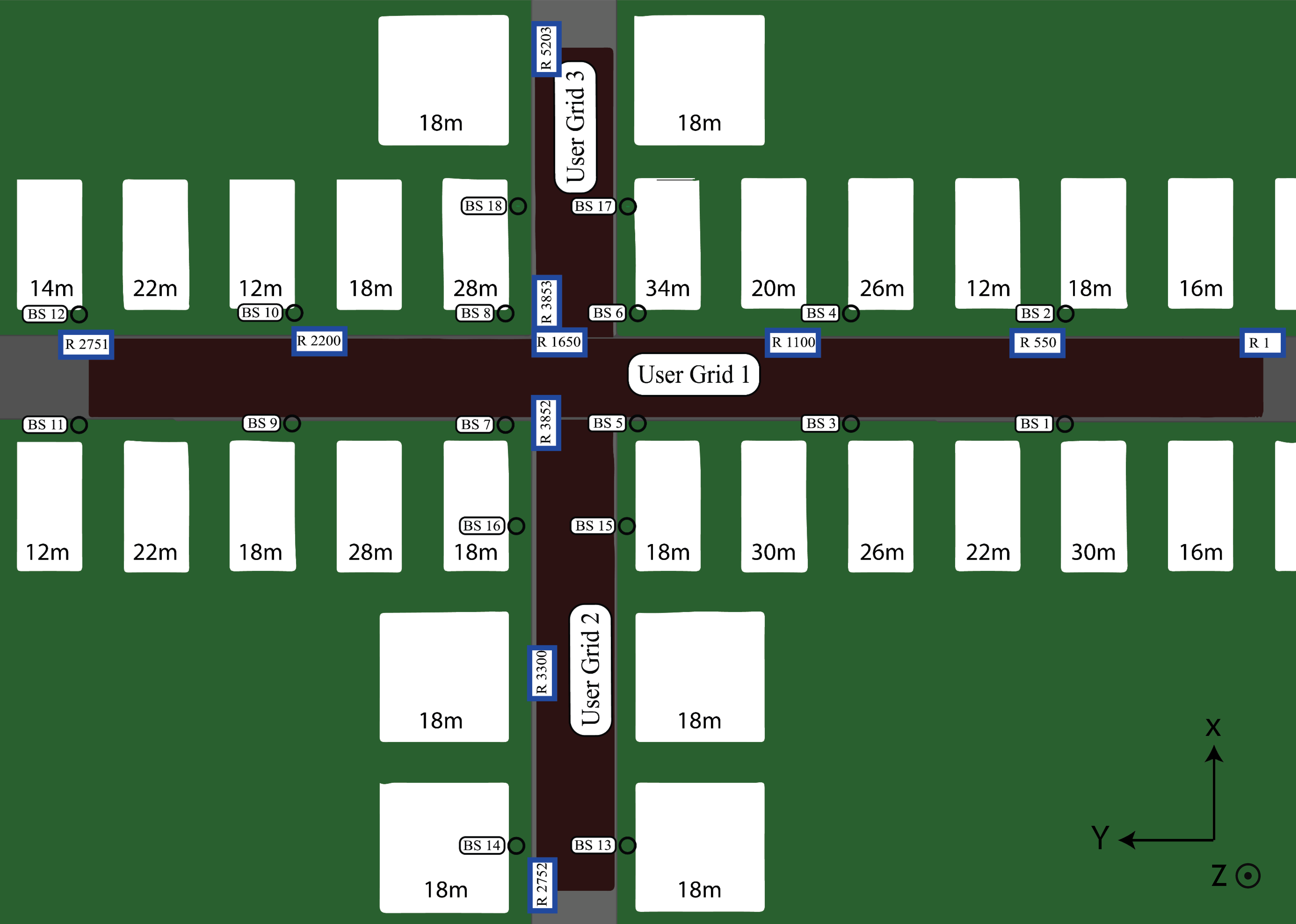}
    \end{center}
    \caption{\; "O1" an outdoor communication scenario of two streets and one intersection at operating frequencies 60 GHz in DeepMIMO \cite{Alkhateeb2019}.  
    } \label{fig:5}
\end{figure}

We employ the O1 scenario from the DeepMIMO dataset for our simulations\cite{Alkhateeb2019}, as shown in \autoref{fig:5}. In the source domain, the BS-3, BS-4, BS-5, and BS-6 serve the users, while in the target domain, users are served by the BS-7, BS-8, BS-10, and BS-12. Each BS has a uniform planar array (UPA) which consists of 32 columns and 8 rows antenna. The antenna spacing is set to half-wavelength, with a system bandwidth of 0.5 GHz operating at 60 GHz carrier frequency. The user is moving on the street at a speed of 30 kilometers per hour. 
\subsection{Simulation Results}
The effective achievable rate of the proposed deep learning and transfer learning method. $R_{{\rm{eff}}}^{{\rm{DL}}}$ can be obtained by
\begin{equation}
\begin{split}
R_{{\rm{eff}}}^{{\rm{DL}}} = \left( {1 - \frac{{2{T_p}}}{{{T_B}}}} \right)\frac{1}{K}\sum\limits_{k = 1}^K {{{\log }_2}} \left( {1 + \frac{{{{\left| {\sum\limits_{n = 1}^{{N_{BS}}} {{\mathbf{h}}_{k,n}^T{{{\bf{\hat{f} }}}_{{\rm{n,}}opt}}} } \right|}^2}}}{{{\sigma ^2}}}} \right)
\end{split}
,\end{equation}
where ${T_p}$ represents beam training pilot sequence time, $2{T_p}$ means the time spent for the uplink pilot sequence and predict optimal beam. Besides, ${{\bf{\hat{f}}}_{n,{\rm{opt}}}}$ denotes the predicted optimal beam at the $n$-th BS that can attain most achievable rate. 

\begin{figure}[tp]
\setlength{\abovecaptionskip}{-0.1cm}
\setlength{\belowcaptionskip}{-0.3cm}
    \begin{center}
    \includegraphics[width=7cm]{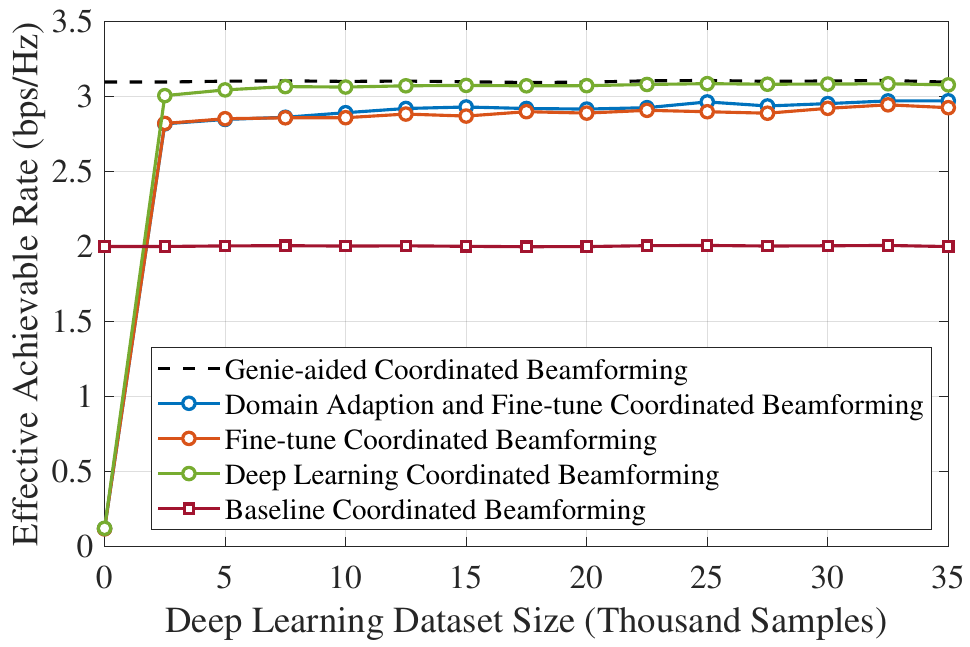}
    \end{center}
    \caption{\; Effective achievable rate of the proposed transfer learning approach
compared to the pure fine-tuning approach and the deep learning approach from scratch.} \label{fig:6}
\end{figure}

Effective achievable rate of different approaches can be shown in \autoref{fig:6}.
Genie-aided coordinated beamforming refers to the scenario with perfect knowledge of the optimal beams as upper bound, while baseline coordinated beamforming denotes exhaustive search over all beams in the codebook. For the pure fine-tuning transfer learning approach, we freeze the  four layers and add two fully connected layers as fine-tuning layers, this approach ensures comparable neural network complexity to the proposed transfer learning-based beam prediction method, while the fine-tuning layers actually contain more trainable parameters than the domain classifier.

The proposed transfer learning-based beam prediction method reaches higher achievable rates compared to conventional fine-tuning approaches. This is because domain adaptation can effectively extract invariant features in the data distributions of the source and target domains, thereby significantly improving the model's performance in the target domain.

\begin{figure}[tp]
\setlength{\abovecaptionskip}{-0.1cm}
\setlength{\belowcaptionskip}{-0.3cm}
    \begin{center}
    \includegraphics[width=6cm]{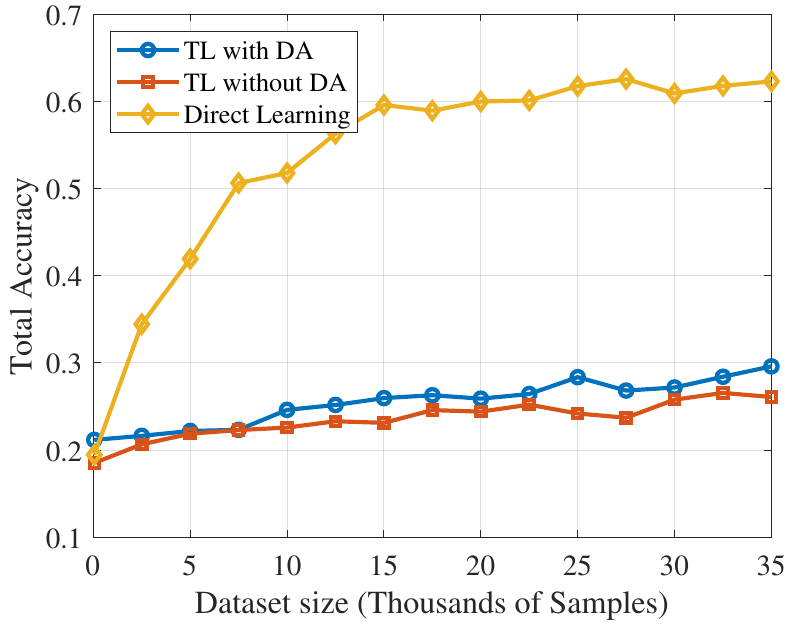}
    \end{center}
    \caption{\; The optimal beam prediction accuracy of different approaches.} \label{fig:7}
\end{figure}

\begin{figure}[tp]
\setlength{\abovecaptionskip}{-0.1cm}
\setlength{\belowcaptionskip}{-0.3cm}
    \begin{center}
    \includegraphics[width=6cm]{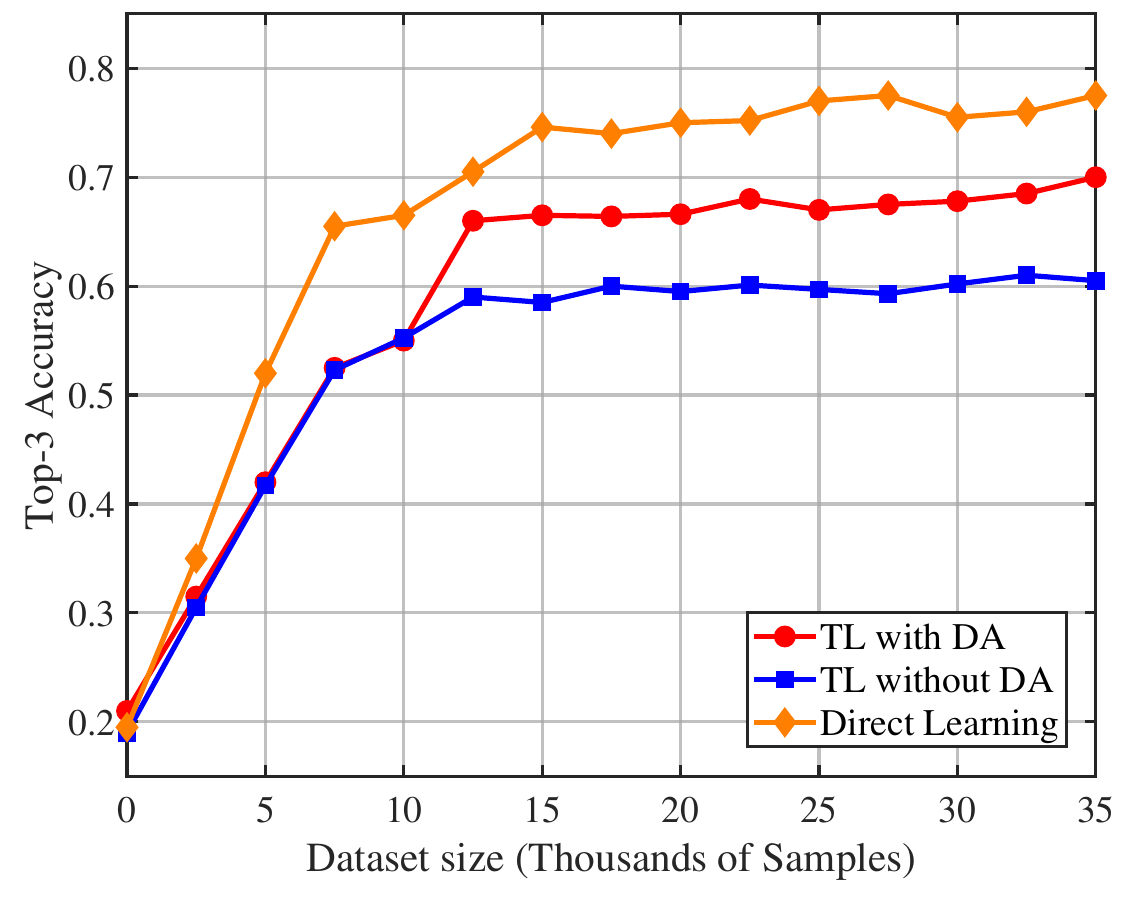}
    \end{center}
    \caption{\; The top-3 beam prediction accuracy of different approaches.} \label{fig:8}
\end{figure}

The optimal beam prediction accuracy for all BSs of different approaches can be shown in \autoref{fig:7}. The proposed transfer learning approach gains higher accuracy compared to the traditional pure fine-tuning approach. Although the accuracy of predicting the optimal beam is not high, the proposed model predicts the achievable rates for all beams in the codebook and selects those with relatively high achievable rates, so the final performance can approach the optimal case. \autoref{fig:8} demonstrates the top-3 beam prediction accuracy across different schemes. \autoref{fig:8} reveals that the proposed transfer learning approach outperforms conventional pure fine-tuning approaches, while achieving comparable accuracy to the method of training the CNN from scratch in the target domain.

\begin{figure}[t!]
    \centering
    \includegraphics[width=5.5cm]{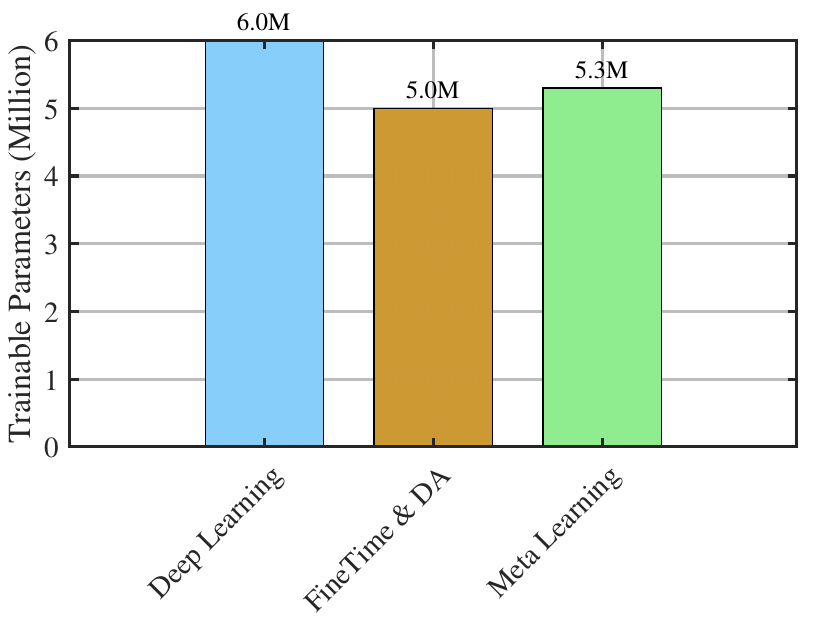}
    \caption{The amount of trainable parameters of different methods.}
    \label{fig:9}
\end{figure}

\autoref{fig:9} shows the total number of trainable parameters in both the conventional CNN and the proposed transfer learning-based beam prediction method. Fine-tune \& domain adaptation (DA) is the proposed transfer learning-based beam prediction method, while Deep Learning is the method of training the CNN from scratch in the target domain. It can be observed that the proposed transfer learning method requires fewer trainable parameters than the traditional meta-learning method and the method of training the CNN from scratch. The proposed method significantly reduces the number of parameters requiring training, thereby decreasing computational overhead.

\section{Conclusion}
In this paper, we proposed a transfer learning-based beamforming vector prediction method for vehicle communication using domain adaptation and fine-tuning methods, which employs a model pre-trained in the source domain. Simulation results in the target domain demonstrate that, building upon previous pure fine-tuning transfer learning approach, the proposed method reaches higher accuracy and a better effective achievable rate. Compared to training the CNN from scratch, the proposed method can reach a comparable achievable rate with fewer trainable parameters, reducing the training overhead.

\bibliographystyle{IEEEtran}%
\bibliography{bibfile}

\end{document}